\documentclass[twocolumn,tighten]{aastex63}
\usepackage{graphicx}
\usepackage{gensymb}
\usepackage{amsmath}

\begin{document}


\title{Discovery of an Exceptional Optical Nebulosity in the Suspected Galactic SN Iax Remnant Pa 30 Linked to the Historical Guest Star of 1181 CE}


\author[0000-0003-3829-2056]{Robert A.\ Fesen}
\affiliation{6127 Wilder Lab, Department of Physics and Astronomy, Dartmouth College, Hanover, New Hampshire, 03755, USA}

\author[0000-0002-2659-8763]{Bradley E.\ Schaefer}
\affiliation{Department of Physics and Astronomy, Louisiana State University, Baton Rouge, Louisiana, 70820, USA}

\author[0000-0002-9018-9606]{Dana Patchick}
\affiliation{Deep Sky Hunters Consortium, Los Angeles, California, 90025, USA}

\begin{abstract}

A newly recognized young Galactic SN remnant, Pa~30 (G123.1+4.6), centered on a hot central star with a $\sim$16,000 km s$^{-1}$ wind velocity has recently been proposed to be the 
result of a double-degenerate merger leading to a SN~Iax event associated with the guest star of 1181 CE. Here we present deep optical [\ion{S}{2}]  $\lambda\lambda$6716,6731 images of Pa~30 which reveal an extraordinary and highly structured nebula $170''$ in diameter with dozens of long ($5'' - 20''$) radially aligned filaments with a convergence point near the hot central star. Optical spectra of filaments indicate a peak expansion velocity $\simeq$1100 km s$^{-1}$ with electron densities of $\leq$100 to 700 cm$^{-3}$, and a thick shell-like structure resembling its appearance in 22 $\mu$m WISE images. No H$\alpha$ emission was seen ([\ion{S}{2}] $\lambda$6716/H$\alpha$ $>5$), with the only other line emission detected being faint [\ion{Ar}{3}] $\lambda$7136 suggesting a S, Ar-rich but H-poor remnant. The nebula's angular size, estimated 2.3 kpc distance, and 1100 km s$^{-1}$ expansion velocity are consistent with an explosion date around 1181 CE. The remnant's unusual appearance may be due to the photoionization of wind-driven ejecta due to clump-wind interactions caused by the central star's high-luminosity wind.

\end{abstract}
\bigskip


\section{Introduction}

For nearly 50 years, the young Galactic supernova remnant (SNR) 3C~58 
has been viewed as the likely remnant associated with the
historical guest star reported by Chinese and Japanese astronomers in early August of 1181 CE, which had a peak brightness $\sim$ 0 mag and  was visible for about six months
\citep{Stephenson1971,Clark1978,Stephenson2002}. 
The location of 3C~58 in Cassiopeia is close to the reported position of the 1181 guest star, and this along with its $\simeq 10^{3}$ km s$^{-1}$ expansion velocity \citep{Fesen2008}, its young 65.7 ms pulsar (PSR J0205+6449) \citep{Murray2002,Camilo2002} plus the lack of any other young SNR known near the guest star's reported location made it the presumptive SN~1181 remnant.

\begin{figure*}[t]
\begin{center}
\includegraphics[angle=0,width=19.0cm]{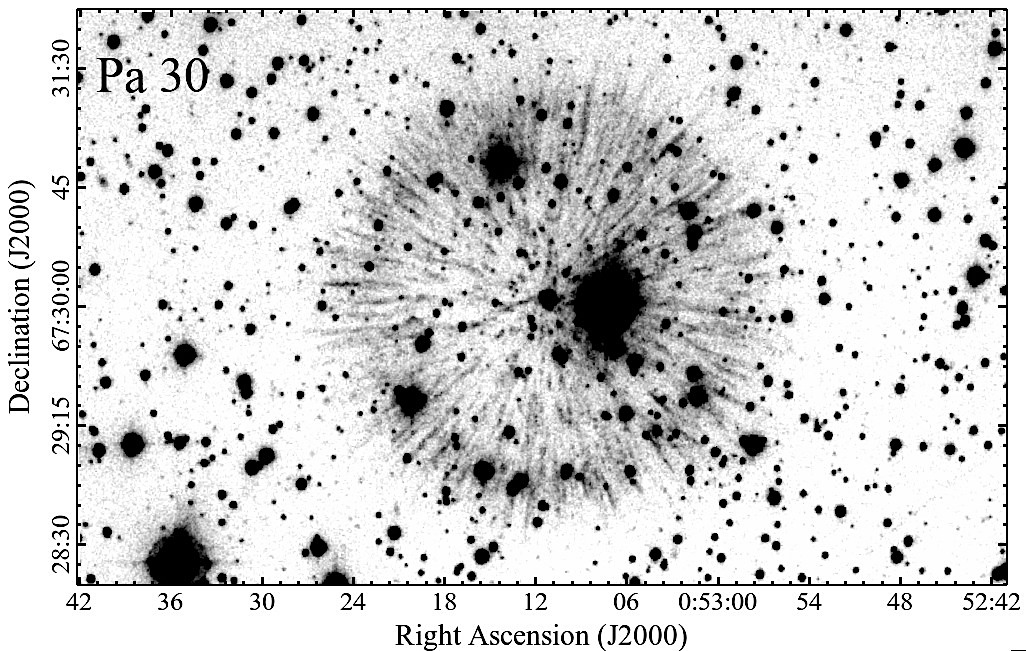}  
\caption{A [\ion{S}{2}] $\lambda\lambda$6716, 6731 image of Pa~30 revealing a highly filamentary and radial morphology. \label{fig1}
}
\end{center}
\end{figure*}

However, this SN -- SNR link was not without problems \citep{Kothes2013,Kothes2017}. Given 3C~58's apparent angular size, proper motion measurements using optical and radio images spanning one or more decades
showed positional shifts far less than
expected if 3C~58 were associated with the historic guest star of 1181 CE \citep{vandenBergh1990,Bietenholz2006,Fesen2008}. 
Moreover, X-ray studies 
set strong upper limits on the thermal emission
from the surface of 3C~58's 66 ms pulsar which fall well below predictions of standard neutron star cooling, thereby
requiring an exotic neutron star cooling processes if linked to SN 1181 and thus only $\sim850$ yr old.

A newly recognized Galactic nebula, Pa~30 (G123.1+4.6), located not far from 3C~58 and actually nearer the spot cited by Chinese and Japanese observers, has recently been identified as the likely true remnant of SN~1181 \citep{Ritter2021,Lykou2022,Schaefer2023}. Initially identified as a possible planetary nebula (Patchick candidate number 30), it was subsequently found associated with a very hot star ($\sim$210,000 K) with an astounding 16,000 km s$^{-1}$ velocity wind \citep{Gvara2019,Garnavich2020}. A young SNR with a hot surviving compact object suggests that it may be the remnant of a double-degenerate merger leading to a subtype Ia SN known as a SN~Iax \citep{Gvara2019,Kashiyama2019,Oskinova2020,Ritter2021,Lykou2022}.

The Pa~30 remnant is bright in the mid-infrared 
\citep{Gvara2019,Ritter2021} and has been detected in X-rays \citep{Oskinova2020}. But no associated H$\alpha$ emission has been firmly detected \citep{Gvara2019,Lykou2022}.  
Deep [\ion{O}{3}] $\lambda$5007 line emission images
revealed a faint and diffuse emission nebula \citep{Kron2014,Ritter2021}
while spectra show [\ion{S}{2}] $\lambda\lambda$6716, 6731 emission with an expansion velocity $\approx$1100 km s$^{-1}$ \citep{Ritter2021}. Because of its known [\ion{S}{2}] emission and a high estimated interstellar extinction ($E(B-V) \simeq$ 0.88; \citealt{Ritter2021,Lykou2022}), 
we thought it worthwhile to explore the optical structure of this young and nearby suspected SN~Iax remnant through red [\ion{S}{2}] images.


\begin{figure*}[t]
\begin{center}
\includegraphics[angle=0,width=18.8cm]{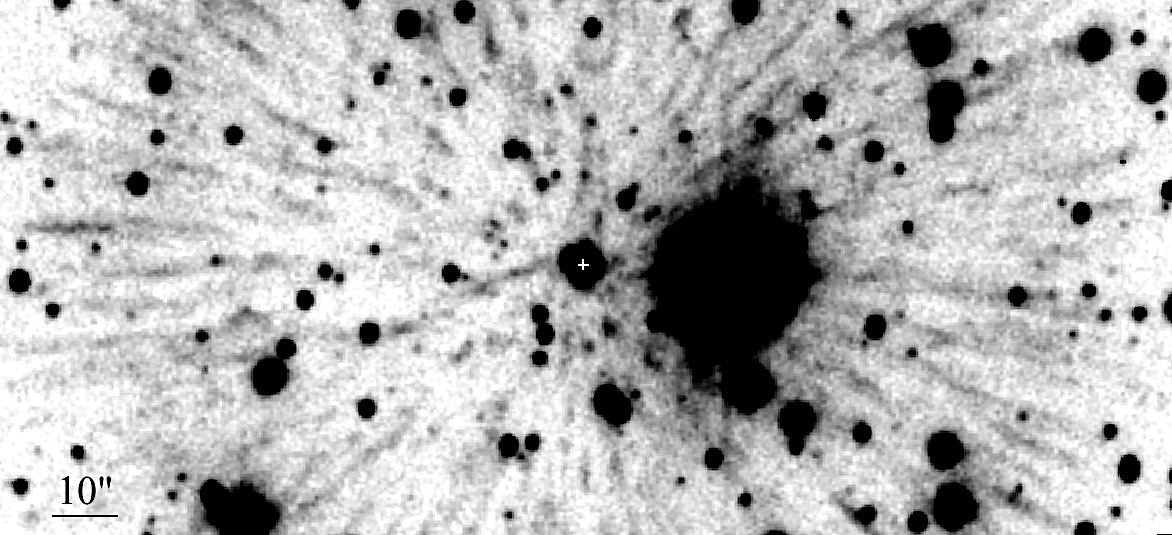} 
\caption{Blowup of Pa~30's central region. A white cross marks the central star, J005311.
North is up, East to the left.
\label{fig2}
}
\end{center}
\end{figure*}


\begin{figure*}[t]
\begin{center}
\includegraphics[angle=0,width=8.5cm]{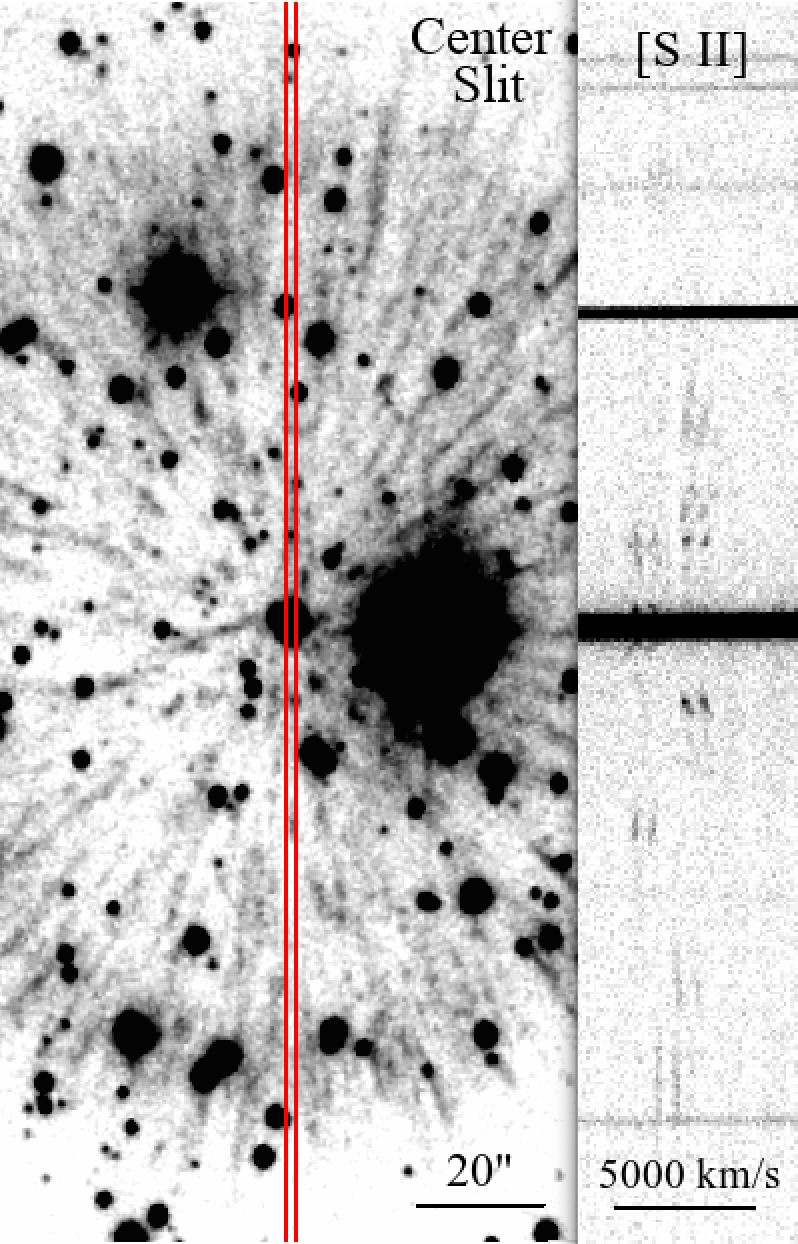} 
\includegraphics[angle=0,width=8.5cm]{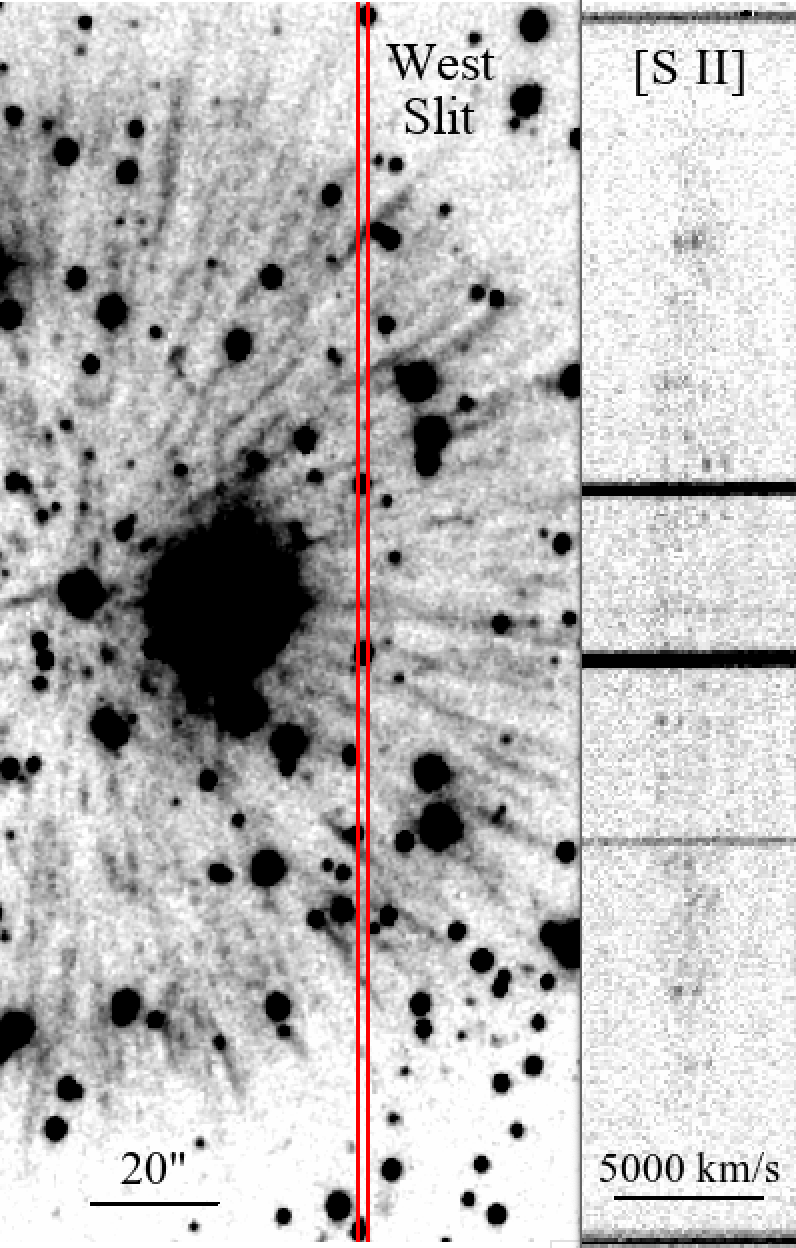} 
\caption{Pair of [\ion{S}{2}] $\lambda\lambda$6716,6731 images of Pa~30 showing central (left) and west (right) slit positions along with the resulting 2D spectra covering the region around the  [\ion{S}{2}] emission doublet with wavelength increasing from left to right. Ordinary background stars have their continuum
spectra visible as horizontal black lines, with the saturated spectrum of Pa~30's
central star, J005311, appearing as an especially wide horizontal black line in the center slit spectrum. 
\label{fig3}
}
\end{center}
\end{figure*}

\begin{figure*}[t]
\begin{center}
\includegraphics[angle=0,width=6.12cm]{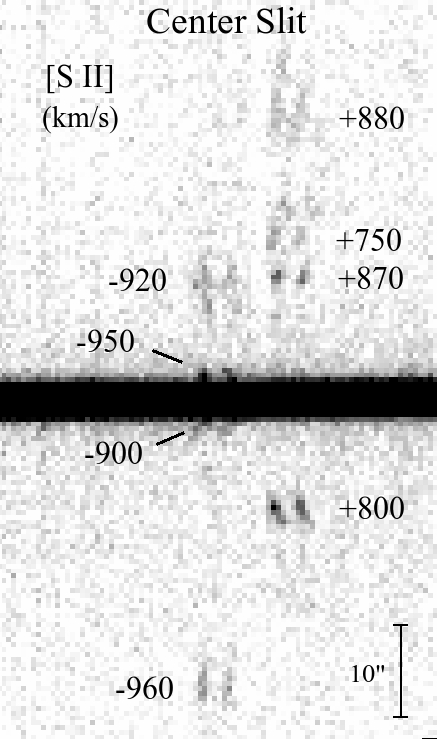}
\hspace{0.5cm}
\includegraphics[angle=0,width=7.12cm]{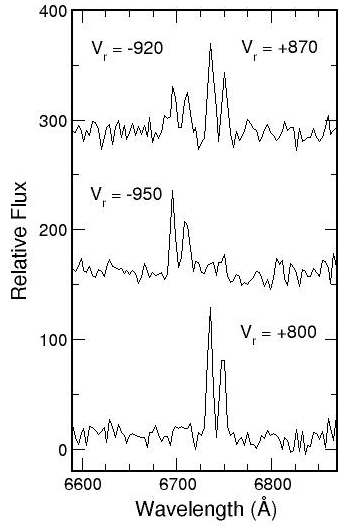}
\caption{{\it{Left:}} Blowup of the middle region of the center slit spectrum showing [\ion{S}{2}] $\lambda\lambda$6716,6731 emissions along with 
measured radial velocities. Wavelength increases from left to right. {\it{Right:}} Spectra for four of the brightest emission knots detected in the center slit covering the [\ion{S}{2}] doublet lines. \label{fig4}
}
\end{center}
\end{figure*}

\begin{figure*}[t]
\centering
\includegraphics[angle=0,width=17.0cm]{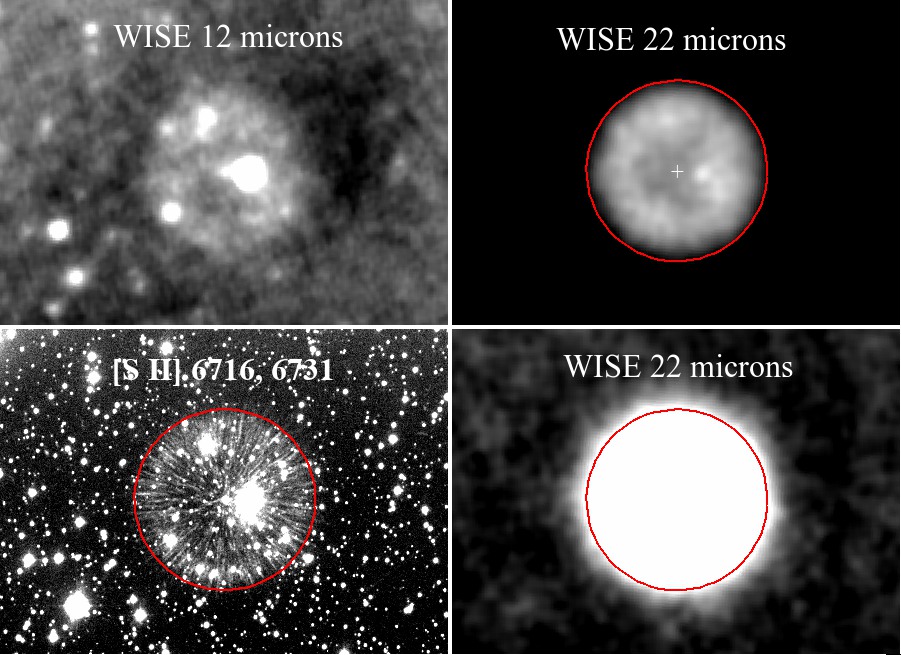}
\caption{ Comparison of WISE 11 $\mu$m    and 22 $\mu$m images with our [\ion{S}{2}] image of Pa~30. Red circles are $85''$ in radius and centered on the hot star, J005311, whose position is marked by a white cross in the upper right panel.
North is up, East to the left. \label{fig5} }
\end{figure*}

 
\section{Observations}

Narrow passband images and low-dispersion spectra of Pa~30 were obtained on 25 and 26 October 2022 with the Hiltner 2.4 m telescope at the MDM Observatory at Kitt Peak, Arizona using
the Ohio State Multi-Object Spectrograph (OSMOS; \citealt{Martini2011}). This instrument employs 
 a $4096 \times 4096$ CCD for both imaging and spectra.  In imaging mode, this telescope/camera system yields a clear FOV of $18' \times 18'$.  On-chip $2 \times 2$  pixel binning gave a spatial resolution scale of $0.55''$ per pixel. Both nights were photometric, with seeing varying from $0.9''$ to $1.3''$.

 Three 2000s exposures were taken of the Pa~30 remnant using a 63 \AA\ FWHM [\ion{S}{2}] 
 $\lambda\lambda$6716, 6731 filter with a T$_{\rm max}$ of 88\% centered at $\lambda$6723. 
 No equally deep off-band images were taken. 
 The [\ion{S}{2}] images were bias subtracted, flat-field corrected, and median filter combined. FK5 WCS coordinates were applied with rms errors less than $0.15''$ in both RA and Dec.
 
Two low-dispersion spectra of the remnant's filaments employing a red VPH grism (R = 1600) and a $15'$ long $1.4''$ wide N-S slit were taken centered on the central star and $\simeq45''$ west of it. Single exposures of 4000s and 4500 s were taken covering the wavelength 4000--8900 \AA \ with a spectral scale of of 3.02 \AA \  pixel$^{-1}$ and a FWHM of 7.8 \AA.
Although spectral coverage was nearly 5000 \AA,  the camera and grism system was most sensitive
in the $6000 - 7500$ \AA\ region. Wavelength calibrations were made through Hg-Ne and Ar comparison lamp spectra. Radial velocity measurements are thought accurate to $\pm$30 km s$^{-1}$ and $\pm$50 km s$^{-1}$ for bright and faint emission features, respectively.


\section{Results}

Figure~\ref{fig1} presents our [\ion{S}{2}] image of Pa~30  which reveals an unexpectedly filamentary remnant unlike any other Galactic SNR. The remnant is 
composed of dozens of long, narrow and radially aligned filaments with typical lengths between $5''$ and $20''$ ($0.06 - 0.24$ pc at d = 2.4 kpc) with length/width ratios $\simeq$ $5 - 10$ given the $\simeq 1''$  image resolution.
Unlike Pa~30's diffuse appearance reported via its faint detection in [\ion{O}{3}] $\lambda$5007  emission \citep{Kron2014,Ritter2021}, there is little diffuse emission seen in our [\ion{S}{2}] image.

The remnant exhibits a very high degree of circular symmetry with a radii 
of $85''$ and $90''$ encompassing $>$95\%  and $>$98\% respectively of the nebula's [\ion{S}{2}] emission. Instead of a sharp and well defined remnant boundary,  its outermost filaments fade rapidly with increasing radial distance past r $\simeq$ $85''$.
While the nebula's filaments are relatively uniform in brightness and length as a function of azimuth angle, one or two contain small knots. Comparison with PanSTARRS r-band images, some stellar-like features near the remnant's center are actually small knot-like emission features. 

The remnant's hot central star, called J005311 based on its J2000 coordinates 
($\alpha$=$00^{\rm h}$53$^{\rm m}$11.2$^{\rm s}$,
 $\delta$=$+67\degr30'2.4''$),
has been proposed to be either a high mass white dwarf (WD), a  super-Chandrasekhar-mass WD, or  the result of a double degenerate merger which resulted in a SN Iax explosion \citep{Gvara2019,Kashiyama2019,Oskinova2020,Lykou2022}.
While the convergence point of the remnant's filaments lies very close to Pa~30's hot central star, some filaments do not appear in perfect alignment with the star's current position. 

This can be seen in Fig.~\ref{fig2} which presents an enlargement of the central region of our [\ion{S}{2}] image. The remnant's hot central star in marked here by a white cross. A relatively bright $12''$ long filament located nearly due east of the central star near the left-hand edge of Fig.~\ref{fig2} is an example of such an apparent mis-alignment, where the filament appears to point back to a position a bit north of the central star current location\footnote{The star's Gaia EDR3 estimated proper motion in declination is $-0.988$ mas yr$^{-1}$ meaning a shift of less than $1''$ in 1000 yr.} 
However, a statistical analysis of an unbiased selection of nearly 100 filaments show a large convergence scatter of some $\pm7''$. This large scatter is likely due to both the image's limited angular resolution and low S/N of many of the remnant's filaments, plus possible confusion caused by overlapping filaments and knots. Consequently, while it is clear that the hot star lies near the center of the [\ion{S}{2}] nebula, we cannot presently determine whether or not if the star's current position is the exact focal point of Pa~30's optical filaments.

Figure~\ref{fig3} shows the precise slit locations along with 2D sky background subtracted spectra for the small  spectral region around the [\ion{S}{2}] doublet for our center and west slit positions. Line emissions from several filaments and knots are readily seen in both slit spectra.
The left-right, right-left tilting of the respective red and blueshifted  [\ion{S}{2}] $\lambda\lambda$6716,6731 lines above and below the spectrum of the center star seen in the center slit spectrum is due to viewing extended emission filaments possessing a velocity dispersion along its length where the velocity increases with distance from the expansion center.

Detected [\ion{S}{2}] emissions in the center slit data are shown  more clearly in Fig.~\ref{fig4}. 
Blueshifted radial velocities ranging from 
$-900$ to $-960$ km s$^{-1}$ and redshifted velocities from $+750$ to $+880$  km s$^{-1}$ are seen close to Pa~30's central star. No low  or intermediate velocity emission features (v$_{\rm r}$ $\leq$ 700 km s$^{-1}$) near the center of the remnant were detected suggesting a shell-like emission structure.

The bright redshifted knot south of the center star with a radial velocity of $+800$ km s$^{-1}$ exhibited a higher velocity `tail' extending to $+990 \pm 50$ km s$^{-1}$. A blueshifted emission feature just below and nearly hidden by the central star showed a velocity dispersion of $-900$ to $-1130$ km s$^{-1}$. The highest velocity of this nearly dead center emission feature plus a
radial velocity of 
$+1060 \pm 50$ km s$^{-1}$ for a faint knot $20''$ north of Pa~30's center star 
suggests a maximum expansion velocity for the nebula of $\simeq$1100 km 
s$^{-1}$, consistent with that reported by \cite{Ritter2021}.

Although fainter, the spectrum obtained in the western portion of Pa~30 looks quite different. It shows knots with a fairly constant red and blue shifts
(v = $\pm 650 \pm100$ km s$^{-1}$) along most of the slit's length until near the north and south ends where the expansion velocities drop below $\sim300$ km s$^{-1}$. 

The electron density sensitive [\ion{S}{2}] $\lambda$6716/$\lambda$6731
line ratio was measured for the brightest filaments and found to range between 1.5 to 0.9 indicating electron densities between $\leq100$ cm$^{-3}$ to $\sim700$ cm$^{-3}$  with an 
average ratio of around 1.3 implying an electron density  $\sim$200 cm$^{-3}$  \citep{Osterbrock2006}. For example, for the four knot spectra plotted in the right panel of Fig.~\ref{fig4}, 
we measure (moving from north to south) I($\lambda$6716)/I($\lambda$6731) values of 1.1, 1.5, 1.3, and 1.3 indicating electron densities ranging from $\leq$100 cm$^{-3}$ to $\simeq300$ cm$^{-3}$.

Although \citet{Ritter2021} do not quote electron densities measured from their GTC/OSIRIS observed [\ion{S}{2}] line ratios, \citet{Lykou2022} states the average [\ion{S}{2}] derived density value from those data was 120 cm$^{-3}$, which is a bit lower than the average we found. Also, in the few cases where there was a significant brightness difference between the ends of a filament, the filaments' brightest portion displayed the lowest velocity, suggesting the densest portions of these filaments lies closer in to the remnant's center.

Besides [\ion{S}{2}] line emission, very weak [\ion{Ar}{3}] $\lambda$7136 emission was also detected but only near the location of the brightest [\ion{S}{2}] emitting knots. 
No clumpy H$\alpha$ emission was detected
at any point along either slit extending at least two arcminutes north and south from Pa~30's central star 
($\lambda$6716/H$\alpha$ $>5$), and no other strong, clumpy line emissions
were seen within the $6000 - 8000$ \AA \ wavelength range 
including [\ion{N}{2}] $\lambda\lambda$6548,6583 
and [\ion{O}{2}] $\lambda\lambda$7319,7330 suggesting a S, Ar rich but H poor emission remnant. However, \citet{Lykou2022} claim the possible detection of faint H$\alpha$ emission from an ``outer Pa~30 nebula'' based upon
a single enhanced pixel at Pa 30's position in the
Virginia Tech Spectral-line Survey (VTSS; \citealt{Dennison1998}). 

Finally, in  the remnant's eastern half, the majority of emission lies within a projected shell of thickness $\sim45''$ starting at an inner radius of $35'' \pm 10''$.  This may not be true for the remnant's western half where the saturated image of an unrelated bright star prevents an accurate assessment. We note that a shell-like structure would be similar in size and shape to the shell-like morphology seen in the WISE 22 $\mu$m image (see Fig.~\ref{fig5}).


\section{Discussion}

Our image and spectral data show Pa~30 to possess a unique and 
unexpected optical nebula, one 
quite unlike any other Galactic supernova remnant (SNR).
Instead of the diffuse remnant reportedly seen in a series of 1200 s [\ion{O}{3}] exposures 
taken in 2013 by G. Jacoby using the KPNO 2.1m \citep{Kron2014,Ritter2021}, when viewed in [\ion{S}{2}] emission Pa~30 displays a nebulosity consisting of dozens of radial aligned filaments pointing back to the center of the nebula.

Why Pa~30 is so different from other young Galactic SNRs may be due in part to the presence of its highly luminous and unusual central star which has a spectrum similar to O-rich WO type Wolf-Rayet stars but with no helium emission. Its \ion{O}{6} line widths indicate an exceptional wind velocity of $16,000 \pm 1000$ km s$^{-1}$,  some three fold higher than seen in any WO type star \citep{Gvara2019,Garnavich2020}.

This high speed wind is not thought to be driven by radiation pressure but rather through magnetic torque and pressure
gradient from the star's strong magnetic field ($\sim$ 10$^{6-8}$ G) and fast spin \citep{Gvara2019,Kashiyama2019,Lykou2022}.
The star also has a high estimated temperature around $210,000$ K, a mass loss rate
of $3.5 \times 10^{-6}$ M$_{\odot}$ yr$^{-1}$ and a log (L/L$_{\odot}$) = 4.6 \citep{Gvara2019,Lykou2022}. Consequently, the Pa~30 SNR is different from most SNRs in that the ejecta we see may not be shock heated but rather photoionized by its central star.

Because of its unusual central star, 
Pa~30 has been suggested as being a SNR arising from a double-degenerate merger leading to a type Ia supernova known as a SN~Iax event \citep{Gvara2019,Kashiyama2019,Oskinova2020,Ritter2021,Lykou2022}.  
SN Iax are characterized as subluminous SNe,  the faintest of which have M$_{\rm V, peak}$ = $-13$ to $-16$ \citep{Sriv2022}, show unusually low expansion velocities near maximum ($2000 - 7000$ km s$^{-1}$; \citealt{Foley2013,Jha2017}) and some may leave a luminous stellar remnant (e.g., SN~2008ha; \citealt{Foley2014}, SN~2012Z; \citealt{McCully2022}). Some SN~Iax such as SN~2002cx, the prototype of the subtype, also show unusually low emission line velocities at late times (700 km s$^{-1}$; \citealt{Jha2006}). However, Pa~30's highly spherical morphology is in sharp contrast to the strongly asymmetric SN~Iax ejecta in some models \citep{Kashyap2018}.

\subsection{The Expansion Velocity of Pa~30} 

Despite its explosive appearance, Pa~30 displays an unusually low expansion 
velocity compared to that commonly seen in SNe or young SNRs.
Based on its optical [\ion{S}{2}] line emissions, Pa~30 has
an expansion velocity of $\sim$1100 km s$^{-1}$ from both on GTC/OSIRIS spectra \citep{Ritter2021} and in our spectral data.
What is unclear, however,
is whether the velocity of its [\ion{S}{2}] filaments might have been  
decelerated  or essentially
undecelerated due to its expansion into a very low density medium around the SN progenitor system. 

The timescale for an ejecta knot deceleration (drag) is given 
by $\tau_{drag} \sim  \chi R_{k}$/V$_{k}$ 
where $\chi$ is the density contrast between the
knot and the ambient medium (i.e., $\rho_{k}$/$\rho_{a}$), R$_{k}$ is the ejecta knot
radius, and V$_{k}$ is the knot’s velocity  which in this case would be 1100 km s$^{-1}$ \citep{Jones1994}.  Because of the unresolved appearance of the handful of clumps in Pa~30's filaments, the dimensions of possible ejecta knots are presently unknown. Choosing an angular size of $0.1''$ corresponds to $\sim 4 \times 10^{10}$ km at a distance of 2.4 kpc. Adopting a low ambient density of the binary WD system prior to SN outburst of 0.01 cm$^{-3}$ hence a $\chi \sim 3 \times 10^{4}$ based on our average electron density of 300 cm$^{-3}$, leads to a $\tau_{drag}$
$\sim 3 \times  10^{4}$ yr, suggesting little appreciable deceleration in 1000 yr. However, much smaller ejecta fragments will decelerate more rapidly owing to their larger surface area-to-mass ratios.

\subsection{The Nature of Pa~30's Long Emission Filaments}

The morphology of Pa~30's filamentary nebulosity does not have close analogues in the remnants of either young Galactic SNe or novae.
Faint trailing ablated material ($\sim 10^{16}$ cm) has been seen for $8,000 - 12,000$ km s$^{-1}$ SN ejecta in the young SNR Cassiopeia A \citep{Fesen2011}. 
Comet-like structures with lengths up to $\sim 10^{17}$ cm off the $600 - 1000$ km s$^{-1}$  nova ejecta clumps seen in 
the case of GK Per (Nova Per 1901)
has been interpreted as  either  ejecta running into a pre-existing ambient medium  \citep{Shara2012} or ablation from clump-wind interactions with dwarf nova wind velocities of several thousand km s$^{-1}$ \citep{Harvey2015}.
Possible analogues to Pa~30's filaments might be the ablated wind blown tails seen in nova DQ~Her \citep{Vaytet2007} which have terminal velocities of  800-900 km s$^{-1}$.

Cometary tails have also been seen in some planetary nebulae, such as the Helix Nebula \citep{Odell1996,Meaburn1998} which have been attributed to a moderately supersonic stream or wind of photoionized gas overrunning a slower expanding mass partially evaporated by photoionization which is then accelerated by momentum transfer \citep{Dyson2006,Matsuura2009}. 
Shadows behind dense clumps immersed in a photoionized medium has also been proposed to explain thin, luminous filaments seen in some PNe \citep{Van1972,Canto1998,Raga2009}.
Long and wide tails have also been seen in bow shocks ``fingers"  behind high-velocity clumps such as in the BN/KL Orion OMC1 outflow \citep{Bally2015}. 

In all these examples there is an identifiable knot or clump undergoing the visible mass loss.  But few if any of Pa~30's 
[\ion{S}{2}] filaments
show evidence of an emission clump or knot at either the start or end of the filaments in our image (cf. Figs. 1 \& 2). Instead, most filaments show little brightness variation along the majority of their $\sim 5 \times 10^{17}$ cm lengths. In addition, the electron density of the Pa~30's filaments is a modest 100-300 cm$^{-3}$, low for SN ejecta which is typically $10^{3-4}$ cm$^{-3}$.

Nonetheless, we view it likely that Pa~30's unusual morphology may be related to the stellar winds of its central star, an object unlike anything in any other nova or young Galactic SNR.
Hydrodynamic models show that supersonic flows produce short tails, while subsonic ones form long tails \citep{Pittard2005,Dyson2006}. This suggests that Pa~30's long filaments are the result of subsonic or mildly supersonic wind off its central star that is accelerating some of the lower density SN ejecta. In such a scenario, an expanding cloud of partially clumpy SN ejecta is sculptured into filaments by Rayleigh-Taylor instabilities through interaction with the central star's stellar winds, then made visible by photoionization by the central star.

Our center slit spectrum supports this picture by showing that some of the [\ion{S}{2}] emission filaments possess a dispersion of velocities of order a few hundred km s$^{-1}$. In addition, the extended emission lines seen in the filament's red and blue shifted velocity dispersion is consistent with the velocity of a filament's emission increasing along its length with increasing distance from the remnant's center much like that seen in clump-wind interactions in novae (DQ Her; \citealt{Vaytet2007}).

Moreover, \citet{Oskinova2020} has suggested that the `hole' seen in the WISE 22 $\mu$m image (see top right panel in Fig.~\ref{fig4}) is due to the  snowplow effect of Pa~30's central star's wind luminosity.
A hole in the remnant's expansion structure would help explain the lack of low and intermediate velocity ejecta (i.e., v$_{\rm r}$ $\leq$ 600 km s$^{-1}$) seen in our center and west slit position spectra (Figs.~3 and 4).
Furthermore, in the SN~Iax merger models of \citet{Shen2017}, the stellar winds off the bound merger WD  may have been considerably stronger in the past, especially during the first one or two decades after outburst.

Although a photoionized clump-wind interaction scenario is highly suggestive, 
we cannot reach a firm conclusion regarding the nature of the Pa~30 filaments given the limited data in hand. Deeper and higher resolution images taken in [\ion{S}{2}] and other emission lines including higher-ionization lines could help clarify the nature of these filaments and the possible presence of filament knots. Higher resolution images may also help determine the nebula's precise convergence point which could then address the possibility of a WD merger star kick as suggested in some SN~Iax models \citep{Kashyap2018,Lach2022}.

\subsection{Pa 30 as the Remnant of SN 1181 CE}

There is now considerable circumstantial evidence linking the Pa~30 SNR to the historical guest star of 1181 CE which might have been a subluminous SN Iax event. Pa~30 lies in a region of the sky close to the recorded position of the historical guest star \citep{Ritter2021,Schaefer2023} and its expansion velocity suggests 
an age consistent with the late 12$^{\rm th}$ century supernova of 1181 CE.

Using estimates of the remnant's energy and the local ISM density, \citet{Oskinova2020} suggested an age of 300-1100 yr. Adopting a remnant radius of $100''$ based on the remnant's 22 $\mu$m WISE data, a distance of $2.3 \pm 0.14$ kpc based on Gaia EDR3 parallax values \citep{Lykou2022} and an undecelerated expansion velocity of $\simeq$1100 km s$^{-1}$ based on GTC/OSIRIS [\ion{S}{2}] line emission, \citet{Ritter2021} estimated an age of $990^{+280}_{-220}$ yr. 

However, using an angular radius of $85''$ more representative of the remnant's [\ion{S}{2}] filaments (see Fig. 5)  yields a current age for Pa~30 of $844 \pm55$ yr. This is in excellent 
agreement with the current 841 yr age of SN 1181.
Adopting instead a 2.4 kpc distance \citep{Schaefer2023} changes this to 
$882\pm 58$ yr, still a very good  match.
These estimates are valid even if the nebula's filamentary emission seen has been wind accelerated as long as the measured ejecta expansion velocity reflects any wind-driven acceleration. 
Consequently, this strongly suggests, but does not prove, that Pa~30 is the remnant of SN~1181 CE.

Finally, the peak visual magnitude of SN~1181 has been estimated to be between 
$+1.0$ and $-0.5$ \citep{Ritter2021} or 0.0 and -1.4 \citep{Schaefer2023} which, assuming 
an A$_{\rm V}$ $\sim$ 2.4-2.9 mag  \citep{Ritter2021,Lykou2022},
implies M$_{\rm V}$ values between $-12$ and $-16$.
Such numbers indicate a subluminous SN event, consistent with the picture of a SN~1181 being a subluminous type Iax event which
shows similar values \citep{Jha2017,Sriv2022}. The presence of a hot central stellar remnant completes the picture of Pa~30 as a young SNR from a double-degenerate merger in a SN~Iax event as some have proposed \citep{Kashiyama2019,Oskinova2020}.

\section{Conclusions}

Our deep [\ion{S}{2}] $\lambda\lambda$6716,6731 images and optical low-dispersion spectra of Pa~30 reveal:

1) A highly structured, radially aligned filamentary nebula with a convergence point near its hot and peculiar central star. 

2) A peak expansion velocity $\simeq$1100 km s$^{-1}$, a complete lack of  low or intermediate velocity filaments, electron densities of $\leq$100 to 700 cm$^{-3}$, and a thick shell-like structure resembling its appearance in 22 $\mu$m WISE images. 

3) No H$\alpha$ emission was detected from the nebula ([\ion{S}{2}] $\lambda$6716/H$\alpha$ $>5$), with the only other line emission detected being very faint [\ion{Ar}{3}] $\lambda$7136 suggesting a S, Ar-rich but H-poor remnant. 

4) The remnant's unusual appearance may be due to the photoionization of wind blown ejecta clump-wind interaction due to the central star's high-luminosity winds.

5) Consistent with the conclusions of \citet{Ritter2021}, \citet{Lykou2022} and \citet{Schaefer2023}, we find that Pa~30 is the likely remnant of the Galactic SN sighted in 1181 CE.  

Higher resolution, multi-wavelength optical and infrared images should help clarify this young remnant's structure, the convergence point of its filaments, and the role of the nebula's central star may have played in the formation of its filaments and central IR cavity. 
Pa~30's relatively small  distance,
its interesting and complex optical and infrared structures centered on a highly unusual star, plus a precisely known age if linked to the supernova event of 1181 CE, all combine to make it an important nebulosity for further study.  \\


\bigskip

We thank the anonymous referee, 
Bruce Balick, Roger Chevalier, 
George Jacoby, Dan Milisavljevic, Dan Patnaude, and Noam Soker for helpful comments and suggestions.
We also thank Eric Galayda and the MDM staff for making these 
observations possible so soon after the June 2022 fire on Kitt Peak mountain. 
This work is part of R.A.F’s Archangel III Research Program 
at Dartmouth. We have made use of the Simbad database, 
NASA's Skyview online data archives, and data from the European Space Agency
mission Gaia (https://www.cosmos.esa.int/gaia), processed
by the Gaia Data Processing and Analysis Consortium (DPAC,
https://www.cosmos.esa.int/web/gaia/dpac/consortium).

\facilities{MDM Observatory (OSMOS)}

\software{PYRAF \cite{Green2012}, XMGrace}

\bibliography{ref2.bib}
\end{document}